# Fundamental Limits to Moore's Law

**Suhas Kumar**

*Department of Electrical Engineering, Stanford University, Stanford, CA 94305, USA.*
*Email: su1@alumni.stanford.edu*

*17 Nov, 2015*

*The theoretical and practical aspects of the fundamental, ultimate, physical limits to scaling, or Moore's law, is presented.*

Gordon Moore had predicted in 1965, that electronic device dimensions will scale following a trend. [1] It is accepted today as the Moore's Law, a rule of thumb that the number of transistors packed in an Integrated Circuit doubles approximately every 2 years. People innovate to stay ahead of the lot by making devices smaller for several advantages, including packing density. [2]

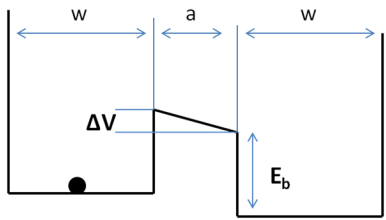

**Fig. 1:** Model for a quantum well computation system.

"Devices" today largely refer to CMOS transistors. "Feature size" today refers to transistor parameters like gate insulator thickness, channel length, etc. or circuit parameters like distance between closest interconnects. Any future electronic device, transistor or not, will see it's limits in the laws discussed here.

## 1. Thermodynamic limits

To perform useful computation, we need to irreversibly change distinguishable states of memory cell(s). The thermodynamic entropy to change n memory cells within m states is $\Delta S = k_B ln(m^n)$, where $k_B$ is the Boltzmann constant. From the second law of thermodynamics, $\Delta S = \Delta Q/T$, where $\Delta Q$ is the energy spent and $T$ is the temperature. So the energy required to write information into one binary memory bit is $E_{bit} = k_B T \, ln2$. This is known as the Shannon-von Neumann-Landauer (SNL) expression. This tells us that we need at least 0.017 eV of energy to process a bit at 300 °K.

From Heisenberg's Uncertainty Principle $\Delta E \Delta t \geq \hbar$, for $E_{bit}$ of 0.017eV, the time to switch is atleast 0.04 ps. From $\Delta x \Delta p \geq \hbar$ or $\Delta x \geq \hbar/\sqrt{2mE}$, the minimum feature size corresponding to an electron as the carrier is 1.5 nm. The power per area, $P = n \times E_{bit}/t_{min}$, where $n = 1/x_{min}^2$ is the packing density (~4.7 $\times 10^{13}$ devices/cm$^2$), is about 3.7 MW/cm$^2$ (The surface of the Sun is 6000 W/cm$^2$). These are not the limits, yet. In the next section, we will correct these formulas by considering tunneling.

## 2. Inclusion of Quantum Tunneling

Consider a quantum well system as shown in Fig 1. The probability of thermionic injection of the electron over the barrier height is $G_T = exp(-E_b/k_B T)$. The probability of tunneling through the barrier is $G_Q = exp(-2a\sqrt{2mE}/\hbar)$. [3] For the two states to be distinguishable, the limiting case is $G_{error} = G_T + G_Q - G_T G_Q = 0.5$. Solving, we get $E_b^{min} = k_B T \, ln2 + (\hbar \, ln2)^2/(8ma^2)$. The power for an area A having n devices operating at a frequency $f$ is $P_{max} = f(n/A)[k_B T \, ln2 + (\hbar \, ln2)^2/(8ma^2)]$.

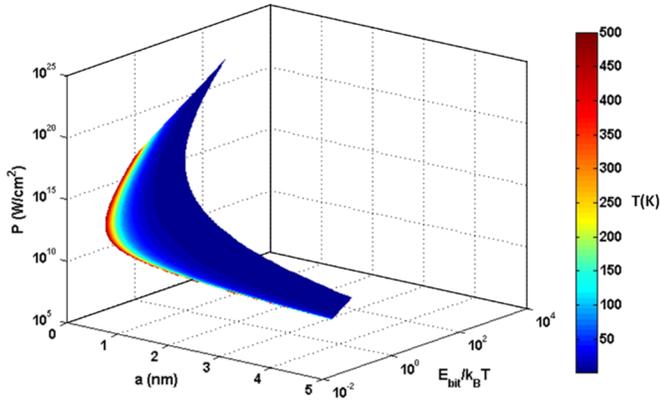

**Fig. 2:** Plot of energy, $E_{bit}$, and power, $P$, as a function of feature size, $a$, at different temperatures.

## 3. Thermal limits

How much we can allow the power to rise depends on how much rise in temperature the chip can stand (typically upto 400 °K) and on how fast we can remove the heat from the chip. Newton's Law of Cooling governs heat removal as $Q = H(T_{Dev} - T_{sink})$. H is the heat transfer coefficient, which is determined by the material constants like specific heat, viscosity, thermal conductivity, heat capacity, etc., apart from the geometry of the cooling structure. [4-5]

When $T_{Dev} < T_{sink}$, it appears from the first section that $E_{bit}$ gets better. But Carnot's theorem says that the work needed to remove heat $Q$ is $W = Q(T_{sink} - T_{Dev})/T_{Dev}$. So

$$E_{bit}^{total} = E_{bit} + E_{bit}(T_{sink} - T_{Dev})/T_{Dev}$$
$$= k_B T_{sink} \ln 2 + (T_{sink}/T_{Dev})(\hbar \ln 2)^2/(8ma^2)$$

$E_{bit}^{total}$ and power are plotted against $a$ and temperature in Fig. 2.

From Fig. 2, we see that (1) cooling the system does not help $E_{bit}$ at all, (2) power is ridiculously high for features less than a nanometer and (3) $E_{bit}$ required is also very high below 2 nm, while it is about $k_B T \ln 2$ for higher features. Notice that, as we approach smaller features, $E_{bit}$ and power are far better behaved at higher temperatures than at lower temperatures.

## 4. Compton wavelength

The Compton wavelength $\lambda_c = h/mc$ (~ 0.00243 nm for $m = 9.1 \times 10^{-31}$ kg) is the characteristic dimension of an electron, which has been proposed as a fundamental absolute limit of the size of an electronic device. [6] At these length scales, there is a run-away-like divergence in power and $E_{bit}$, as apparent from Fig. 2. The reader is encouraged to plug in this length scale into the equations presented here (and into Fig. 2) to estimate power and $E_{bit}$ and decide if it is sensible to even approach this limit.

## 5. Practical aspects

Power consumption and speed are limited fundamentally by the devices, but practically by the electrical parasites, interconnects and chip architecture. This is the reason for the clock speed to have saturated at about 3 GHz for today's processors. All alternative ideas, like optical interconnects and more would see their limit in the domain conversion, which is limited by thermodynamics discussed here. The "2" in $k_B T \ln 2$ can be made higher to, say, $m$, but that only pushes the limits by a factor of $\ln(m)/\ln 2$. [7-8]

## 6. Conclusions

We will hit these scaling limits in 30-40 years. We do not know if we can compute in other methods that are governed by laws yet to be explored. For example, we are not sure if we will realize fundamental locally active components like memcapacitor and meminductor, which remain fantastic predictions for computing without power. [9]